\documentclass[11pt,a4paper]{article}
\usepackage{jcappub}

\newcommand{\beq}{\begin{equation}}
\newcommand{\eeq}{\end{equation}}
\newcommand{\beqa}{\begin{eqnarray}}
\newcommand{\eeqa}{\end{eqnarray}}

\begin{document}
\title{Dark Matter Mini-halo around the Compact Objects: the Formation, Evolution and Possible Contribution to the Cosmic Ray Electrons/Positrons}
\author[a,b,c]{Rui-Zhi Yang}\emailAdd{bixian85@pmo.ac.cn}
\author[a,b]{Yi-Zhong Fan}%\emailAdd{yzfan@pmo.ac.cn}
\author[d]{Roni Waldman}%\emailAdd{waldman@cc.huji.ac.il}
\author[a,b]{Jin Chang}%\emailAdd{chang@pmo.ac.cn}
\affiliation[a]{Purple Mountain Observatory, Chinese Academy of Sciences,\\ Nanjing
210008, China}
\affiliation[b]{Key laboratory of Dark Matter and Space Astronomy, Chinese Academy of Sciences,\\ Nanjing
210008, China}
\affiliation[c]{Graduate University of Chinese Academy of Sciences,\\ Beijing, 100012, China}
\affiliation[d]{Racah Institute of Physics, Hebrew University, Jerusalem 91904, Israel}

\abstract{Dark matter particles may be captured by a star and then thermalized in the star's core. At the end
of its life a massive star collapses suddenly and a compact object is formed. The dark matter particles
redistribute accordingly. In the inelastic dark matter model, an extended dense dark matter
mini-halo surrounding the neutron star may be formed. Such mini-halos may be common in the Galaxy. The electron/positron flux resulting in the annihilation of dark matter particles, however, is unable to give rise to observable signal unless a nascent mini-halo  is within a distance $\sim ~{\rm a~few}~0.1$ pc from the Earth.}

\maketitle
\flushbottom
\section{Introduction}
Dark matter (DM) is a form of matter necessary to account for gravitational effects observed in very large scale structures such as anomalies in the rotation of galaxies and the gravitational lensing of light by galaxy clusters that cannot be accounted for by the quantity of observed matter \cite[e.g.,][]{lsp,lkp}.
The existence of dark matter is a piece of solid evidence for new physics that is beyond the standard model and the relevant researches are the focus of modern physics and astrophysics.
Among the hypothetical particles proposed so far, the weakly interacting massive particles (WIMPs) are the leading
candidate \cite{lsp,lkp}. The direct-detection searches look for
the signal of WIMP-nuclei scattering in underground detectors. The indirect-detection
experiments, instead, aim to catch the annihilation/decay products of DM particles.
Currently there are some anomalous signals in favor of WIMPs but the evidence is not conclusive \cite{atic,pamela,fermi,fan}.

The DM particles may be captured by the stars. The capture of WIMPs by the sun,
earth and other astrophysical objects has been extensively investigated \cite{gould1,gould2,press85}.
 The captured particles will eventually be thermalized and centered in the
core of the star, namely, a DM core forms in the center of
a star \cite{peter47}. After the death of a (massive) star, a compact object is formed. The collapse of the star lasts for a very short time and the density distribution of the DM core will be modified. If the compact object is small enough,
a significant part of the DM core may be left outside the compact object, i.e.,
a DM mini-halo is formed. As the outer shell of the stellar remnant eventually disappears, the annihilation products of the DM matter, such as electrons and photons, are visible for the observers. In this work we investigate the formation and evolution of such a mini-halo and calculate the flux of the annihilated electrons/positrons. This work is structured as follows. In section 2 we describe the distribution of
WIMPs inside the star. In section 3 we investigate the redistribution of WIMPs in the collapse process.
In section 4 we calculate the evolution of the DM mini-halo and in section 5 we estimate the electron/positron
flux detectable on the earth. We summarize our results in section 6.

\section{Dark matter capture by the star}
The WIMPs scatter with nuclei and lose their kinetic energy when
pass through a star. These particles will get captured if the kinetic energy is
smaller than the stellar gravity potential. After being captured, the WIMPs will be
eventually thermalized in the star \cite{peter47}. The density distribution of the WIMPs is
described by \cite{nus09,ther}
\begin{equation}
\rho(r)=\rho_{\rm center} e^{-m_{\rm \chi}\phi(r)/T}=\rho_{\rm
center} e^{-r^2/r_{\rm th}^2},
\end{equation}\\
where $T$ is the temperature of the center of the star, $\phi(r)$ is the gravitational potential, $m_\chi$ is the
rest mass of WIMPs, $\rho_{\rm center}$
is the dark matter number density in the center of the star and
\begin{equation}
r_{\rm th}=(\frac{3T}{2\pi m_{\rm \chi}G\rho_{\rm center}})^{1/2}=0.01 R_{\odot}
(\frac{T}{1.2\rm keV})^{1/2}(\frac{100\rm GeV}{m_{\rm \chi}})^{1/2}(\frac{150\rm GeV}{\rho_{\rm center}})^{1/2},
\end{equation}
where $R_{\odot}$ is the solar radius. As a result, a DM core is formed and the annihilation efficiency is given by \cite{Fan2011}
\begin{equation}
C_{\rm A}={\int d^3 r \rho^2(r) <\sigma_{\rm A} v> \over (\int d^{3} \rho(r))^{2}}=\frac{<\sigma_{\rm A} v>}{(2\pi)^{3/2}r_{\rm th}^3},
\end{equation}
where $<\sigma_{\rm A} v>\sim 3\times 10^{-26}~{\rm cm^{3}~s^{-1}}$ is the averaged annihilation cross section.
The number of total DM particles ($N$) evolves as
\begin{equation}
\frac{d N}{d t}=C-C_{\rm A} N^2,
\end{equation}
where $C$ is the capture rate and the annihilation rate can be
expressed as $\Gamma _{\rm A} =\frac{1}{2}C_{\rm A} N^2$. Eq.(2.4) can be solved analytically and we have
\begin{equation}
\Gamma_{\rm A}=\frac{1}{2}C {\rm tanh^2}(t/\tau _{\rm eq}),
\label{eq:5}
\end{equation}
where $\tau _{\rm  eq}=1/(CC_{\rm A})^{1/2}= 10^7 \rm yr (\frac{r_{\rm th}}{10^9 \rm cm})^{2/3} (\frac{10^{-26} \rm cm^3 s^{-1}}{<\sigma_{\rm A} v>})^{1/2} (\frac{10^{25} \rm s^{-1}}{C})^{1/2}$. If $\tau _{\rm eq}$ is
longer than the lifetime of the star, the equilibrium can never be
reached. If equilibrium is reached the annihilation
rate is $\Gamma _{\rm A} =\frac{1}{2}C$.

In most works, the
scatter is assumed to be elastic. However, the recent results from
DAMA and CDMS may favor the inelastic DM
model \cite{dama,dama2,dama3}. In such a model, the WIMPs
will scatter inelastically with the nuclei so the kinetics is
modified \cite{in1,in2,in3}. The cross
section in the inelastic case can be a few orders of magnitude larger than that in the elastic
case. Nussinov et al.
\cite{nus09} discussed the inelastic capture of DM particles
by the sun and found that the
annihilation-capture processes can reach equilibrium but the density profile of the DM core is very different
from that formed in the elastic capture. For example, the radius of the DM core is much large. Some DM particles go outside the star \cite{out}.

\section{Dark matter distribution after the stellar collapse}
In the calculation and simulation below, we take DM mass $m_{\rm \chi} = 100 \rm GeV$, elastic scatter cross section $10^{-44}\rm cm^2$, inelastic scatter cross section $10^{-41}\rm cm^2$ and DM halo energy density $0.3\rm GeV/cm^3$ as the fiducial values.

\subsection{DM distribution and Stellar Evolution}
For our purpose we need to trace the stellar evolution and then the DM distribution inside the evolving star. We consider two examples, a $4 M_{\odot}$ star as a typical progenitor of white dwarfs, and a $15 M_{\odot}$ star as a typical progenitor of a neutron star, both at solar metallicity. The evolution of both models is followed with the public code MESA \citep{MESA}, using the standard numerical methods and input physics as described therein. The $4 M_{\odot}$ model has been evolved to the formation of a white dwarf, using the mass loss formula for AGB stars by \cite{blocker}, while the evolution of the $15 M_{\odot}$ model was followed to the formation of the iron core using the mass loss formula of \cite{vink} for hot stars and of \cite{dejager} for cool stars. The density profiles of these stars at the end of their lives are shown in Fig.1 and Fig.2, respectively.

The DM distribution, without doubt, changes with the evolution of the stars. The process can be approximated by the adiabatic contraction process.  The response of individual WIMP to the adiabatic contraction of the star can be traced with the adiabatic invariants $J$ and $J_{\rm r}  = 2\int_{r_{\rm min}}^{r_{\rm max}}\sqrt{2[E-\psi(r)]-J^2/r^2} dr$, where $J$ is the angular momentum and $J_{\rm r}$ is the radial action. $E$ is the total mechanical energy of the particle and $\psi(r)$ is the gravitational potential at radius $r$. These invariants are not modified when the star change its gravitational potential in the evolution if the particles do not scatter. So we can generate the distribution of the WIMPs in every stage of the star evolution with these invariants. The detailed description of such a process can be found in \cite{AC} and we follow their approach in this work.
\\

On the other hand the DM particles are captured continuously, which lose their energy by further scattering and their density profile also changed due to the adiabatic contraction. Both effects, the scattering and the adiabatic contraction, are considered in our following investigation.
\begin{figure}
\includegraphics[width=150mm,angle=0]{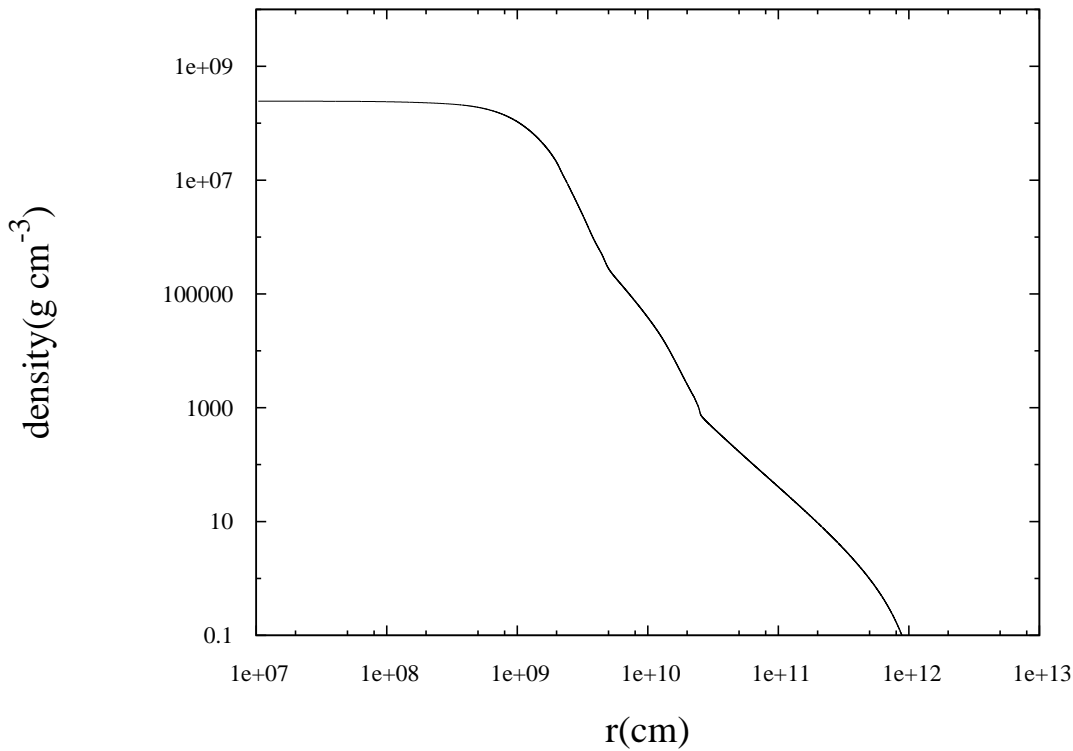}
\caption{The number density profile of the star with an initial mass $15 M_{\odot}$ just before its collapse.
}
\end{figure}

\begin{figure}
\includegraphics[width=150mm,angle=0]{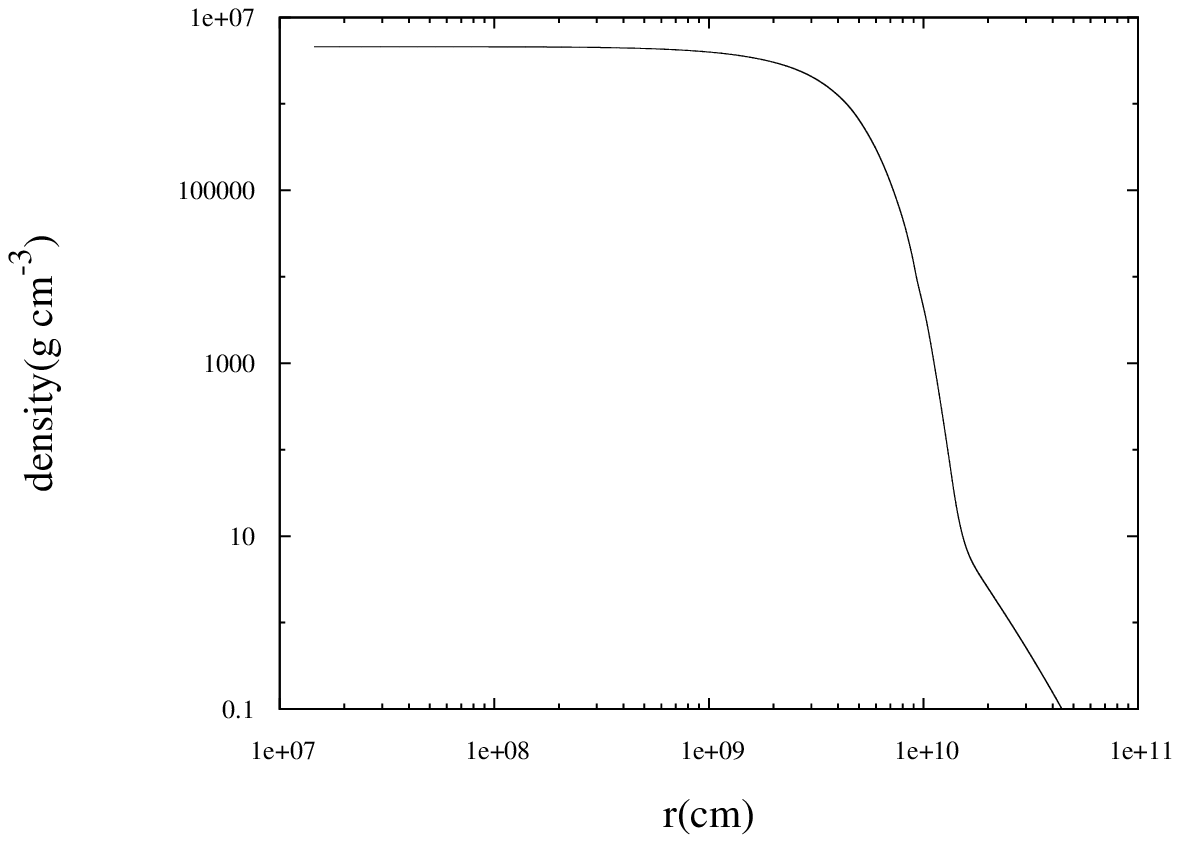}
\caption{The number density profile of the star with an initial mass $4 M_{\odot}$ just before the formation of a white dwarf.}
\end{figure}

\subsection{Elastic case}
%In the elastic case the density distribution of the thermalized DM particles are described in the last section.
From kinematics we know that a WIMP will lose about $m_{\rm nu}/m_{\chi}$ of its kinetic energy in each elastic scatter. To thermalize a WIMP needs to lose almost all the kinetic energy, i.e., The WIMP
will need to scatter roughly
 $m_{\chi}/m_{\rm nu}$ times. Note that the time needed for one scatter is $\frac{1}{\rho_{\rm nu}\sigma_{\rm s} v}$. Then the thermalization timescale can be estimated by
\begin{equation}
\tau_{\rm ther}=\frac{m_{\chi}/m_{\rm nu}}{\rho_{\rm nu}\sigma_{\rm s} v}
\end{equation}
where $m_{\rm nu}$ is the rest mass of the nuclei, $\rho_{\rm nu}$ is the number density of the nuclei in the star, $\sigma_{\rm s}$ is the elastic scatter cross section and $v$ is the velocity of the WIMPs inside the star. We take $v$ to be the average of the escape velocity $v_e$ and the
final velocity $v_f$, where $v_f== (3kT /m_{\chi} )^{1/2}$ assuming the dark matter has
thermalized with the star and has a temperature which is equal to the
temperature of the star at its center 	.
 in the core of the star in our estimation. %Note that $\tau_{\rm ther}$ is the time scale for the DM to be thermalized inside the star,
%which is different from $\tau_{eq}$, the time scale in which the annihilation and capture process become equilibrium.
The central density and the duration of the stars in different burning stages are shown in TABLE.1 in \cite{massive}, with which it is straightforward to estimate $\tau_{\rm ther}$ and then examine whether the thermalization can be established or not. For the stars of $15 M_{\odot}$ and $4 M_{\odot}$ the thermalization can be reached in all these stages. Consequently the DM density in the star can be approximated by an isothermal sphere.
Before collapse, the thermalized WIMPs distribute as \cite{you80}
\begin{equation}
\rho(r)=\frac{\pi^{1/2}}{{2\sqrt{2}}}\int_{\rm
\phi(r)}^{0}\sqrt{2(E-\phi(r))}{\rm exp}[-E/kT]dE,
\end{equation}
where $\phi(r)$ is the gravitational potential before collapse and $E$ is the total mechanical energy of the particle.  $\phi(r)$ and
 temperature $T$ s evaluated at the end of the star's evolutionary track. The baryon density in the core of the star can be taken as a constant, with which the integration of Eq.(3.2) give the Eq.(2.1), where the value $r_{\rm th}$ can now be evaluated with equation (2.2) and star evolution results. The central $T$ is $3.6 \times 10^{9}~{\rm K}$ and $1.3 \times 10^{9}~{\rm K}$ for the $15 M_{\odot}$ star and $4 M_{\odot}$ star, respectively. The density profiles are shown in Fig.1 and Fig.2. Then $r_{\rm th}$ is $10^7 {\rm cm}$ for the $15 M_{\odot}$ and $10^9 ~{\rm cm}$ for the $4 M_{\odot}$ star.
As the star evolves the baryonic profile will adjust. The dark matter profile will
also evolve due to adiabatic contraction.  The distribution
 function in the energy space is needed, which reads \cite{you80}
\begin{equation}
\rho(r,E)=\int_{ J_{\rm max}}^{0}\frac{JdJ f(E)}{r^2
[2(E-\phi(r))-J^2/r^2]^{1/2}},
\end{equation}
\begin{equation}
f(E)=(2\pi)^{-3/2}{\rm exp}(-E),
\end{equation}
where $J$ is the angular momentum, and $J_{\rm max} = r\sqrt{{2[E-\phi(r)]}/{m_{\rm \chi}}}$ is the maximum angular momentum at radius $r$.

For white dwarfs, no further collapse happens and the above result is the final profile of WIMPs if the annihilation can be ignored. But for neutron stars, the collapse makes the situation more complicated.  After collapse the gravitational potential has
changed a lot, it is thus very interesting to see whether a
significant fraction of WIMPs is left outside the compact object or not. The change of the distribution of WIMPs in the collapse process of the medium with a constant density has been discussed in \cite{cocollapse}. But in reality the density profile of the dying stars is much more complicated (see Fig.1). So we adopt a numerical approach. We treat the collapse as a free fall process and calculate the gravity in each time step of the collapse, then calculate the corresponding energy change of each dark matter particle, until the stellar core has
formed a neutron star which has a radius of $12 \rm km $. After the collapse presumably, an explosion happens, which sweeps away material outside of
the nascent $1.5 M_{\odot}$ neutron star.
 The distribution of WIMPs before collapse is simulated by the use of MonteCarlo method with eq.(3.2) and eq.(3.3).
We are interested in the WIMPs surrounding the compact
object. Since the compact object is very
dense, any WIMPs passing through will scatter with the nuclei so
frequently that lose their kinetic energy, and finally sink into
the center of the compact object. That's why we drop the WIMPs whose semi-minor axis is smaller than the radius of the compact object.  At the end of the life of the $4~M_\odot$ star, $r_{\rm th} \sim 10^{9}~{\rm cm}$ is comparable to the typical radius of white dwarfs, so most WIMPs are concentrated in the white dwarf and are not of our interest. On the other hand, at the end of the life of the $15~M_\odot$ star, $r_{\rm th} \sim 10^{7}~{\rm cm}$ is only one order larger than the neutron star's radius and nearly all the WIMPs falls inside the neutron star in the collapse in our simulation. Hence no prominent dark matter mini-halo can be formed.

\subsection{Inelastic case}
The inelastic case \cite{in1,in2,in3,smith01} is different from the elastic case in some
aspects. For example, an inelastic scatter will not occur if the energy transfer is not large enough. The energy transfer
in a scatter is \cite{press85}
\begin{equation}
\Delta {\cal E} = \frac{2m_{\rm \chi}m_{\rm nu}}{(m_{\rm \chi}+m_{\rm nu})^2}(1-
\cos \theta){\cal E},
\label{eq:energy_transfer}
\end{equation}
where ${\cal E}$ is the WIMP's kinetic energy and $\theta$ is the scatter angle. It should be mentioned that $m_{\rm nu}\sim 50~{\rm GeV}$ since among all nuclei the iron has the greatest contribution to the total cross section \cite{in1}. In deriving eq.(\ref{eq:energy_transfer})
 the nucleus is assumed to be at rest \cite{peter47}. Inelastic scattering does not happen if
${\cal E}$ is below a threshold ${\cal E}_{\rm th}$ given by $\Delta {\cal E} = \delta$, which reads
\begin{equation}
{\cal E}_{\rm th}=(1+\frac{m_{\rm \chi}}{m_{\rm nu}})\delta,
\end{equation}
where $\delta$ is the mass gap in the inelastic DM model.
For $m_{\chi}\sim 100$ GeV and $\delta\sim 200$ keV that are consistent with all current experiments \cite{dama,dama2} we have ${\cal E}_{\rm th} \sim~\rm 600 keV$.

Generally speaking, the inelastic dark matter should also scatter elastically due to elastic coupling \cite{nus09} or loop corrections if there is no tree level elastic coupling\cite{batell09}. But the cross section of such elastic scattering is highly model-dependent. \cite{nus09} suggested that the elastic cross section can be six orders of magnitude smaller than the inelastic cross section. It is shown in \cite{batell09} that the cross section can be as large as $10^{-42}\rm cm^2$ if the force mediator is very light while it can also be as small as $10^{-50}\rm cm^2$ if the force mediator is heaver than $1 \rm GeV$. If the elastic cross section is large enough the dark matter inside the star is thermalized, for which  our discussion in last subsection applies and there is no novel phenomena taking place. However if the elastic cross section is as small as $10^{-50}~{\rm cm^2}$ the thermalization can not be reached in our star evolution model. Such a scenario is the focus of the following discussion.
With our Eq.(3.1) and Table 1 of \cite{massive} it is straightforward to show that the thermalization of dark matter particles is established most easily in the helium burning stage. Therefore we have a threshold of the elastic cross section below which the dark matter particles cannot be thermalized in all burning phases. For our $15 M_{\odot}$ model the threshold is $\sim 10^{-46}~{\rm cm^2}$ while for our $4 M_{\odot}$ model the threshold is $\sim 10^{-47}~{\rm cm^2}$.  As a result, if the elastic cross section is very small, say, about $10^{-50}~\rm cm^2$ as mentioned above, thermalization cannot be reached. We assume it is the case in our following discussion.
 %To simplify the discussion and in order to close any possible loophole in the argument we neglect the effects due to elastic scattering in the inelastic dark matter model.

Due to the inelasticity, WIMPs do not thermalize with the star and do not have a thermal distribution as in elastic
case. A WIMP will lose most of its kinetic energy after sufficient scattering and then it has a
residual kinetic energy below ${\cal E}_{\rm th}$, which depends mainly on the
kinematics of the last scattering. So the DM density profile can not be treated as an isothermal sphere any longer. A simple assumption is that the kinetic energy of the WIMPs after enough scattering has a flat distribution ranging from zero to $\delta$.
If the possibility of the scattering at different radius is known, the distribution of total energy $E$ can be derived by adding the kinetic energy and gravitational potential. But note that the possibility of scattering at different radius depends on the particle's orbits, which will change significantly in former scatters. So in this work we carry out a Monte-Carlo simulation to trace the particles' fate inside the star.
%Every particles will be treated individually.
We calculate their orbits after each scattering and then simulate the next scattering with the kinematics described above. The evolution of stars of $15 M_{\odot}$ and $4 M_{\odot}$ has been divided into tens of intervals. We calculate the DM capture rate and the resulting DM density profile in each interval and also take into account possible scatter and adiabatic contraction in the later phases.
The rest of the simulation, i.e., the DM density change due to the collapse, is similar to the elastic case.
Please see figure 3 and figure 4 for the results for a neutron star and a white dwarf, respectively. In the case of
a neutron star there are a large amount of WIMPs outside the compact
object, i.e., a DM mini-halo with a radius of $\sim 2\times10^9~{\rm cm}$ is formed. In the case of white dwarf, the radius of the compact object is only a few times smaller than the size of the DM mini-halo. Hence no prominent mini-halo forms surrounding the white dwarfs.

Please note that in the simulation we use the fiducial DM halo density $0.3 \rm GeV/cm^3$. In some regions of the galaxy, such as the galaxy center and some DM clumps, the DM density can be much higher. If the lifetime of the star is much smaller than the equilibrium time $\tau_{\rm eq}$, the annihilation of DM inside the star is not important compared with the capture. So the DM density of the mini-halo is %proportional to the total number of captured DM, i.e.,
proportional to the capture rate and then the local DM density. If before collapse the equilibrium has already been established, the DM density in the mini-halo should be proportional to the equilibrium density. In such a case the capture rate $C$ is equal to the annihilation rate $C_{\rm A} N^2$, which means that the DM density is proportional to the square root of the local DM density. Therefore in both cases the denser the environment DM, the higher the DM density of the mini-halo.
\begin{figure}
\includegraphics[width=150mm,angle=0]{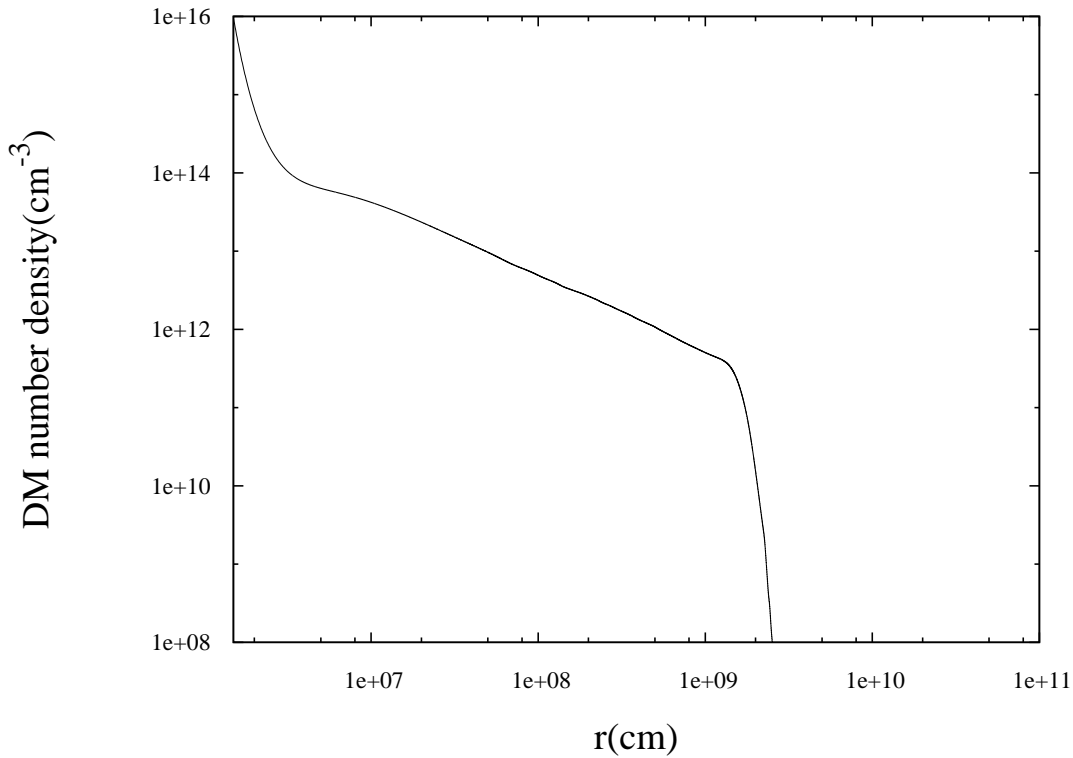}
  \caption{The density distribution of the DM mini-halo surrounding the neutron star: the inelastic DM model.}
\end{figure}
\begin{figure}
\includegraphics[width=150mm,angle=0]{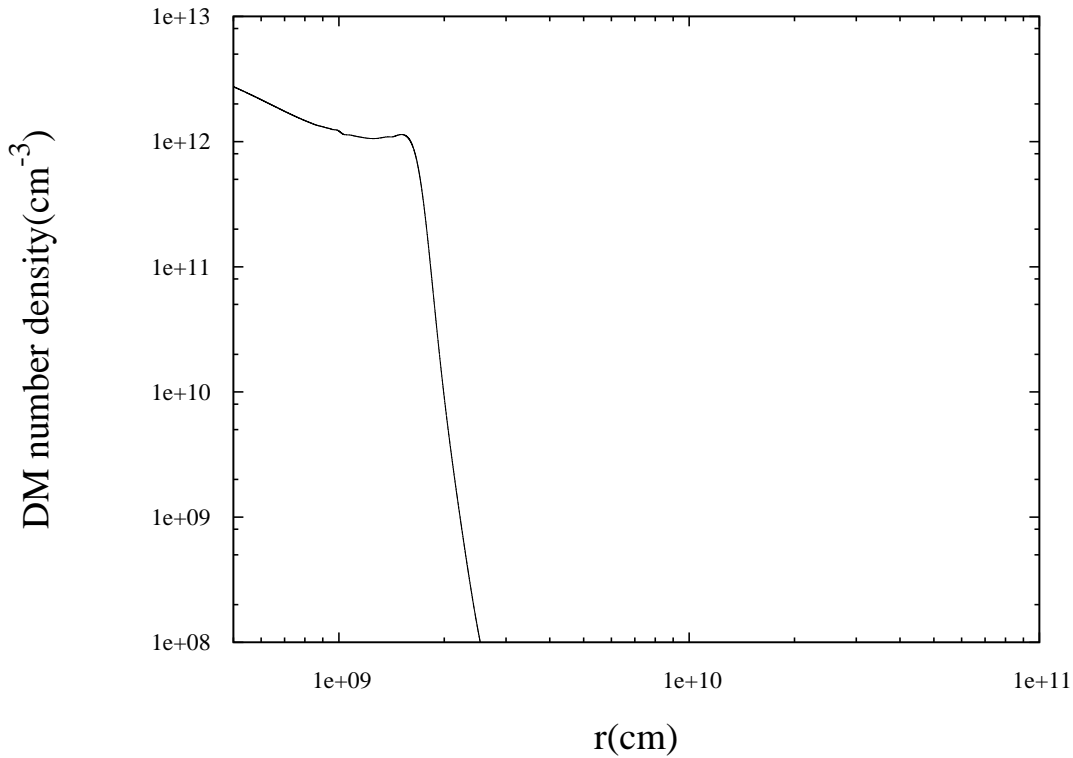}
  \caption{The density distribution of the DM mini-halo surrounding the white dwarf: the inelastic DM model}
\end{figure}

\subsection{General relativity correction}
Near the compact objects, in particular the neutron star, the gravitation
is so strong that general relativity may be not ignorable any longer. In the
simulation we find that the captured particles all have a very large
eccentricity, so most orbits lie far from the central star, where
the general relativity effect can be ignored. The
general relativity effect we are interested in is that the particles moving
around the central star would have no stable orbits if their angular momentum is small enough. As shown in \cite{sha87}, if the condition
\begin{equation}
J<2\sqrt{3}\frac{GMm_{\rm \chi}}{c}
\end{equation}
is met the particles will be captured by the central object in a Schwarzschild metric,
where $M$ is the mass of the compact object and  $c$ is the speed of light. Other less important general relativity effects such as the
distortion of orbit shape and time delay have been ignored.

\subsection{Recoil velocity}
When the star collapses a huge amount of energy is released. The energy release is not always isotropic and the formed compact object may
get a considerable recoil velocity. For the neutron star the typical
velocity is about $300~{ \rm km~s^{-1}}$ and for white dwarf it is
much smaller. Since the collapse is transient, the recoil can be
taken as a `kick' of the particles in the compact object's frame,
i.e., all particles captured by the star get a velocity at the same
direction and their kinetic energy change. When the collapse happens the gravity is so strong that the escape velocity is of the order $10^{4} {\rm km/s}$ , so a small change of the kinetic energy can be safely neglected in our simulation.

\section{Time evolution of the density of the DM mini-halo}
As mentioned above, only in the case of inelastic scattering for a
neutron star does an extended mini-halo form. Below we focus on such a specific scenario.
A WIMP in the DM mini-halo will maintain its orbit except annihilating with the other WIMP or scattering
with baryons. Since we consider only the WIMPs whose orbits are totally outside the neutron star, the scattering with baryon can be neglected. The time evolution of the density of WIMPs is governed by
\begin{equation}
\frac{d \rho}{d t}=-<\sigma_{\rm A} v>\rho^2.
\end{equation}
 The solution is simply
\begin{equation}
 \rho = \frac{1}{1/\rho_{\rm 0}+<\sigma_{\rm A} v> t},
\end{equation}
where $\rho_{\rm 0}$ is the number density at time 0, i.e., the moment
right after the collapse. The supernova remnant surrounding neutron star persists for about 	 $\sim 10^6$
years, after which the remnant material becomes
transparent and the products of WIMPs annihilation become
detectable. For $t \sim 10^6$ years and $<\sigma_{\rm A} v> \sim
10^{-26}~\rm cm ^3 s ^{-1}$,  number density $\rho$ won't change significantly unless it is larger than $10^{13}~\rm
cm^{-3}$.  For the
cosmological timescale $t \sim 10^{10}$ years, the DM number density evolve due to annihilation can be
as high as $\sim 10^{8}~\rm cm^{-3}$. From fig.3 we find that the DM number density in the nascent mini halo is much larger than $\sim 10^{8}~\rm cm^{-3}$.  As a result of the annihilation, the number density drops with time as
 $\rho \sim 10^{10}(t/10^{8}~{\rm year})^{-1}~ \rm cm^{-3}$ for $t\geq 10^{8}$ years.

\section{Flux of annihilation products}

Now we can estimate the flux of the annihilation products of WIMPs in the mini-halo
surrounding the neutron stars. The annihilation rate is
\begin{equation}
N_{\rm anni} = 4\pi \int <\sigma_{\rm A} v> \rho^2 (r)r^2 dr.
\end{equation}
In the inelastic DM model, as shown in fig.3, the WIMPs are centered
in a small region $r\sim  2 \times 10^{9} {\rm cm}$ with the number density $\rho(r)  \sim 10^{12}~{\rm cm^{-3}}$ shortly after the collapse. The current
number density is in order of $10^{8}~{\rm cm^{-3}}$ and the annihilation rate is $\sim 10^{19}~{\rm s^{-1}}$  if the compact objects are old enough.
Let's estimate the contribution of the Galactic DM halo. The local DM energy density is $\rm{0.3~GeV~cm^{-3}}$, for $m_{\rm \chi}\sim \rm 100~{\rm GeV}$
the annihilation rate of Galactic DM halo particles in $\sim {\rm 1~kpc^3}$ space is about $10^{34}$ per
second. The number densities of local neutron stars is
$\sim 2\times 10^6~{\rm{kpc^{-3}}}$
% while for white dwarf it's $1.5\time10^6~{\rm{kpc^{-3}}}$
 \cite{sha87}. So for inelastic
WIMPs the annihilation rate of these mini-halos in $\rm 1~kpc^3$ is about $10^{25}~{\rm s^{-1}}$, i.e., The amount of DM annihilating in the mini-halo is small compared to the Milky Way Halo.

It is interesting to investigate whether the annihilation of inelastic DM particles in some nascent mini-halos can produce enough products giving rise to detectable signal.
To account for the cosmic electron/positron data at an energy $\sim 100$ GeV,
the annihilation rate should be $\sim 10^{36}-10^{37}~{\rm s^{-1}~kpc^{-3}}$. Therefore no observable signal is expected unless one nascent mini-halo is within a distance $\sim {\rm a~ few~0.1~pc}$ to the Earth.

\section{Conclusion}
In this work we have investigated the formation, evolution and the annihilation of the particles in dark matter mini-halo surrounding the compact object. In our model, the dark matter particles are captured by the stellar matter and then distributed in the core of the star. At the end
of its life the star uses up the fuel and a compact object,
either a white dwarf, a neutron star or a black hole,  is formed. The dark matter particles will
redistribute around the newly formed compact object. In the elastic dark matter model and if the elastic cross section is large enough in the inelastic dark matter model, the dark matter particles will be thermalized and no prominent dark matter mini-halo can be formed. In the inelastic dark matter model, however, if the elastic cross section is very small (for example $\lesssim 10^{-47}~{\rm cm^{2}}$) an extended dense dark matter mini-halo surrounding neutron star may be formed. Though such mini-halos may be very common in the Galaxy, the electron/positron flux resulting in the annihilation of these dark matter particles
is negligible compared with the observational data unless a nascent mini-halo is within $\sim$ a few $0.1~{\rm pc}$ to the Earth.
%Please note that such a conclusion is very robust since the capture of the yielded electron/positron by the compact object,
%not discussed in this work, will lower the detectable flux further.

\acknowledgments We thank Dr. Q. Yuan for comments. This work was supported in part by National Basic
Research Program of China under grants 2009CB824800 and 2010CB0032, and by National Natural Science of China under grants 10920101070, 10925315 and 11073057. YZF is also supported by the 100 Talents Program of Chinese Academy of Sciences.

\end{document}